\documentstyle[12pt,aps,prb]{revtex}

\renewcommand{\vec}[1]{{\mathbf #1}}

\begin{document}

\draft

\title{Nonuniversal correlations in multiple scattering}
\author{S. E. Skipetrov and R. Maynard}
\address{Laboratoire de Physique et Mod\'elisation des Milieux
Condens\'es,
Universit\'e Joseph Fourier, Maison des Magist\`eres
--- CNRS,
B.P. 166, 38042 Grenoble Cedex 9, France}

\date{\today}
\maketitle

\vspace{1cm}
\begin{abstract}
We show that intensity of a wave created by a source embedded
inside a three-dimensional disordered medium exhibits
a non-universal space-time correlation
which depends explicitly on the short-distance properties of
disorder, source size, and dynamics of disorder in the
immediate neighborhood of the source.
This correlation has an infinite spatial range
and is long-ranged in time. We suggest that a technique
of ``diffuse microscopy'' might be developed employing
spatially-selective sensitivity of the considered correlation
to the disorder properties.
\end{abstract}


\section{Introduction}
\label{intro}
Tremendous progress has been made during the last years in
the understanding of multiple scattering of waves in disordered media in
so-called mesoscopic regime.\cite{rossum99}
This regime is commonly defined by
\begin{eqnarray}
\lambda \ll \ell \ll L \ll L_{abs},
\label{meso}
\end{eqnarray}
where $\lambda$ is the wavelength, $\ell$ is the mean free
path, $L$ is the sample size (or source-detector separation
in the case of infinite medium), and $L_{abs}$ is the
absorption length. In particular, considerable attention
has been paid to the study of the correlation
function of intensity fluctuations\cite{rossum99,feng88,berk94}
\begin{eqnarray}
C(\vec{r}, \Delta \vec{r}, \tau) = \frac{\left< \delta I(\vec{r}, t)
\delta I(\vec{r}+\Delta \vec{r}, t+\tau) \right>}{\left<
I(\vec{r}, t) \right> \left<I(\vec{r}+\Delta \vec{r}, t) \right>},
\label{corr}
\end{eqnarray}
where $I(\vec{r}, t) = \left| \psi(\vec{r}, t) \right|^2$ is
the intensity of the wave field $\psi(\vec{r}, t)$,
$\delta I(\vec{r}, t) = I(\vec{r}, t) -
\left< I(\vec{r}, t) \right>$, angular brackets $\left< \cdots \right>$
denote ensemble averaging,
and the random fields $\psi(\vec{r}, t)$, $I(\vec{r}, t)$ are
assumed to be stationary.
In the present paper, we are mainly interested in
the correlation of intensities in two distant points
separated by a distance superior to the mean free path:
$\Delta r > \ell$.

Several contributions to $C$
can be identified.\cite{feng88,berk94}
Short-range (in both space and time) correlations arise from
interference between the waves arriving at the points $\vec{r}$
and $\vec{r} + \Delta \vec{r}$. Weak long-range correlations are due
to the interaction of diffusing modes in the medium.
Finally, tiny infinite-range correlations manifest themselves in the
phenomenon of universal conductance fluctuations,\cite{umbach84}
and have been recently observed for optical waves.\cite{frank98}
The listed contributions to the correlation function
are commonly denoted by $C_1$, $C_2$ and $C_3$, respectively.
It is important to emphasize that the magnitudes of
these correlations scale with some dimensionless parameter
and are not sensitive to the short-distance properties of disorder
provided that $\left| \vec{r} - \vec{r}_0 \right|,
\left| \vec{r} + \Delta \vec{r} - \vec{r}_0 \right| > \ell$.
For the case of plane wave transmission through a slab, such a
parameter is the dimensionless conductance $g \sim N \ell /L$,
where $N \sim k_0^2 A$ is the number of transport channels,
$k_0 = 2\pi/\lambda$ is the wavenumber,
and $A$ is the area of the slab surface.
For $g \gg 1$, which is a typical experimental situation,
one finds\cite{feng88}
$C_1 \sim 1$, $C_2 \sim g^{-1}$, $C_3 \sim g^{-2}$.
For a medium of infinite extent, the relevant parameter is
the product $k_0 \ell$: $C_1 \sim 1$, $C_2 \sim (k_0 \ell)^{-2}$,
$C_3 \sim (k_0 \ell)^{-4}$, where $k_0 \ell \gg 1$ is assumed.
In all cases, the only medium-specific quantity
determining the magnitude of the correlation is the
mean free path $\ell$, and thus we call
the $C_1$, $C_2$, and $C_3$ correlations {\em universal\/} as
they are independent of the particular features of disorder,
and only depend on the global parameter $\ell$.

Recently, a new contribution to the intensity correlation in
random media ($C_0$) has been conjectured.\cite{shapiro99} This contribution
originates from the scattering of wave near its source.
In the case of a point source embedded in an infinite three-dimensional
medium with fluctuations of the dielectric constant obeying
white-noise Gaussian statistics, for $\Delta r > \ell$ and
$\tau = 0$, it has been found\cite{shapiro99}
\begin{eqnarray}
C_0(\vec{r}, \Delta \vec{r}, 0) =
\frac{\pi}{k_0 \ell}.
\label{c0}
\end{eqnarray}
The physical process giving rise to the $C_0$ correlation
consists in the scattering of the wave
in the immediate neighborhood of its source (more precisely, at a distance of
order $\lambda$ from the source), which produces a secondary wave
propagating to the distant points $\vec{r}$ and $\vec{r} + \Delta \vec{r}$
by diffusion. It turns out that the secondary waves
created in such a way are correlated, and hence the described physical
process establishes correlation between intensities at
different points of the medium. It is important to
emphasize that this correlation has an infinite spatial range.

A fundamental difference between $C_0$ and other types of
correlation is that $C_0$ should be sensitive to the
short-distance details of disorder (e.g., to its
correlation length).\cite{shapiro99}
Moreover, as this contribution to the correlation function
is due to the scattering in the vicinity of the source, one could
expect that $C_0$ should depend on the geometrical parameters
of the source and, in particular, on its size.
Finally, for moving scatterers and $\tau \neq 0$,
$C_0$ should also be sensitive to the dynamics of
disordered medium in the immediate neighborhood of the
source.
It follows then that the $C_0$ contribution to the correlation
function is {\em non-universal.}

In the present paper, we show that the $C_0$ correlation function does
exhibit a strongly {\em non-universal\/} behavior.
First, we study the dependence of $C_0$ on the correlation length
of disorder $\ell_{\varepsilon}$. It is shown that the $C_0$ correlation
function depends explicitly on $k_0 \ell_{\varepsilon}$, and,
more precisely, that it decreases linearly with
$k_0 \ell_{\varepsilon}$ for $k_0 \ell_{\varepsilon} \ll 1$.
The sensitivity of the $C_0$ correlation function to the correlation
length of disorder might be used for sensitive
specific measurements of $\ell_{\varepsilon}$.
Next, we study the effect of the source size $a$ on the $C_0$
correlation.
We show that $C_0$ decreases as $a$ is increased.
The decrease of $C_0$ with $a$ is linear for $k_0 a \ll 1$ and becomes
power-law [$C_0 \propto (k_0 a)^{-1}$] for $k_0 a \gg 1$.
Finally, we compute the space-time intensity correlation function
$C_0(\vec{r}, \Delta \vec{r}, \tau)$. We show that this correlation function
is sensitive to the motion of scatterers
in the immediate neighborhood of the source. 
In the case of disordered medium consisting of scatterers
undergoing Brownian motion, $C_0$ decays linearly with
$\sqrt{\tau/\tau_0}$
as $\tau \ll \tau_0$ and exhibits a power-law behavior
[$C_0 \propto (\tau/\tau_0)^{-3/2}$] for $\tau \gg \tau_0$, with
$\tau_0$ being a single-scattering correlation time.
The decay of $C_0$ with time is
exclusively determined by the motion of scatterers located at a distance of
order $\lambda$ from the source, and is not sensitive to the dynamics
of the rest of the medium. 
Measurements of $C_0$ could therefore
provide a sensitive spatially-selective probe of scatterer motion
in disordered media.

\section{Spatial correlation}
\label{spatial}

Following Ref.\ \onlinecite{shapiro99}, we consider an infinite
three-dimensional disordered
medium with a monochromatic
source of scalar waves located at $\vec{r}_0$. We assume
that the source is spherically symmetric and has a radius $a$.
To treat the problem of non-zero $a$, one needs to specify the
boundary conditions for the wave field at
the surface $\left| \vec{r} - \vec{r}_0 \right| = a$. We assume that
waves are transmitted through the region of space occupied by
the source without being scattered.\cite{source}
Then for
$\left| \vec{r} - \vec{r}_0 \right| > a$
the wave field $\psi(\vec{r}, t)$ obeys the scalar
wave equation with a point source at $\vec{r}_0$:
\begin{eqnarray}
\left\{ \nabla^2 + k_0^2 \left[1 + \delta \varepsilon(\vec{r}, t)
+ i \eta \right] \right\} \psi(\vec{r}, t) = \delta(\vec{r}-\vec{r}_0),
\label{weq}
\end{eqnarray}
where $\delta \varepsilon(\vec{r}, t)$ is the fluctuating part of the
dielectric constant
[$\delta \varepsilon(\vec{r}, t) \equiv 0$ for
$\left| \vec{r} - \vec{r}_0 \right| < a$],
and $\eta$ is a positive infinitesimal ensuring
the uniqueness of solution of Eq.\ (\ref{weq}).
The wave equation can be written in the form of
Eq.\ (\ref{weq}) only if $\delta \varepsilon(\vec{r}, t)$ does not
change significantly at the time scale of $(c k_0)^{-1}$,
with $c$ being the speed of propagation in the average medium:
$\partial/\partial t [\delta \varepsilon(\vec{r}, t)] \ll c k_0$.

Let us first consider the purely spatial correlation function
$C(\vec{r}, \Delta \vec{r}, 0)$. The time evolution
of $\delta \varepsilon(\vec{r}, t)$ is of no importance in this case,
and the statistical
properties of $\delta \varepsilon(\vec{r}, t)$ are completely specified by
its average value and covariance.
Without loss of generality,
we consider $\delta \varepsilon(\vec{r}, t)$ which obeys
Gaussian statistics with zero average and correlation length
$\ell_{\varepsilon}$:
\begin{eqnarray}
&&\left< \delta \varepsilon(\vec{r}, t) \right> = 0,
\label{zeroaverage}
\\
&&B_{\varepsilon}(\Delta \vec{r}) =
\left< \delta \varepsilon(\vec{r}, t)
\delta \varepsilon(\vec{r}+\Delta \vec{r}, t) \right> =
\sigma_{\varepsilon}^2 \exp\left(
-\frac{\Delta r^2}{\ell_{\varepsilon}^2} \right),
\label{corrfunc}
\end{eqnarray}
where $\left| \vec{r} - \vec{r}_0 \right|,
\left|\vec{r} + \Delta \vec{r} - \vec{r}_0 \right| > a$.
In the present paper we restrict ourselves to
weak scattering
($k_0^2 \ell_{\varepsilon}^2 \sigma_{\varepsilon}^2 \ll 1$)
and short correlation length ($\ell_{\varepsilon} \ll \lambda$).
The first assumption allows us to benefit from a standard
diagrammatic techniques\cite{frisch68} in the first order
Born approximation,
while the second one permits to not distinguish between the
scattering and the transport mean free paths. The case of
$\ell_{\varepsilon} \gtrsim \lambda$ calls for a more sophisticated
treatment which is beyond the scope of this paper.
For $k_0^2 \ell_{\varepsilon}^2 \sigma_{\varepsilon}^2 \ll 1$
and $\ell_{\varepsilon} \ll \lambda$, the mean free
path is $\ell = 4 \pi /(u k_0^4)$ with
$u = \pi^{3/2} \sigma_{\varepsilon}^2 \ell_{\varepsilon}^3$.
In the following we assume that any variation
of $\ell_{\varepsilon}$ implies corresponding change of
$\sigma_{\varepsilon}$ so that $\ell$ remains unchanged.

As shown in Ref.\ \onlinecite{shapiro99},
in a weak scattering limit ($k_0 \ell \gg 1$),
the leading contribution to
$\left< \delta I(\vec{r}, t)
\delta I(\vec{r}+\Delta \vec{r}, t+\tau) \right>$
can be represented by a
diagram depicted in Fig.\ \ref{fig1},
where the ladder propagators are $L(\vec{r}_1, \vec{r}_2) =
3/(\ell^3 \left| \vec{r}_1 - \vec{r}_2 \right|)$, and
$V$ is a vertex to be calculated with account for the short-distance
properties of disorder and the size of the source.
This vertex is a short-range object and can be replaced by a number.
We show the leading contributions to $V$ in Fig.\ \ref{fig2}. Note that
the diagrams (c) and (d) contribute to the vertex only
if $\ell_{\varepsilon} \neq 0$ or $a \neq 0$ 
and therefore they have not been counted in Ref.\ \onlinecite{shapiro99},
where only the case of $\ell_{\varepsilon} = a = 0$ has been
considered.

The calculation of the diagrams of Fig.\ \ref{fig2} is performed
similarly to that of Ref.\ \onlinecite{shapiro99}.
For the diagram (b) of Fig.\ \ref{fig2} we obtain, for instance,
\begin{eqnarray}
V^{(b)} = k_0^4 \left(\frac{\ell}{4 \pi}\right)^2
\int d^3 \vec{r}_1 \int d^3 \vec{r}_4
\overline{G}(\vec{r}_1 - \vec{r}_0)
\overline{G}^*(\vec{r}_4 - \vec{r}_0)
f^*(\vec{r}_1 - \vec{r}_0)
f(\vec{r}_4 - \vec{r}_0)
B_{\varepsilon}(\vec{r}_1, \vec{r}_4),
\label{vb}
\end{eqnarray}
where integrations cover the whole space
except the interior of a sphere of radius $a$ centered at $\vec{r}_0$,
\begin{eqnarray}
\overline{G}(\vec{r}-\vec{r}_0) =
-\frac{1}{4 \pi \left| \vec{r}-\vec{r}_0 \right|}
\exp\left[ \left( i k_0  - \frac{1}{2 \ell} \right)
\left| \vec{r}-\vec{r}_0 \right| \right]
\label{green}
\end{eqnarray}
is the average Green function of Eq.\ (\ref{weq}), and
\begin{eqnarray}
f(\vec{r}-\vec{r}_0) =
\frac{\sin(k_0 \left| \vec{r}-\vec{r}_0 \right|)}{k_0 \left|
\vec{r}-\vec{r}_0 \right|}
\exp\left( - \frac{\left| \vec{r}-\vec{r}_0 \right|}{2 \ell} \right).
\label{f}
\end{eqnarray}
The derivation of the two above expressions
is performed under important assumptions of $k_0 \ell \gg$
and $a \ll \ell$. The second assumption allows us to neglect the
absence of the dielectric constant fluctuations inside
the region of space occupied by the source, which
simplifies greatly the derivations.

Results similar to Eq.\ (\ref{vb}) are also obtained for the three
other
diagrams of Fig.\ \ref{fig2}. The diagram (c), for example, is given by
the same integral as in the Eq.\ (\ref{vb}) with the integrand
$\overline{G}^*(\vec{r}_1 - \vec{r}_0)
\overline{G}^*(\vec{r}_4 - \vec{r}_0)
f(\vec{r}_1 - \vec{r}_0)
f(\vec{r}_4 - \vec{r}_0)
B_{\varepsilon}(\vec{r}_1, \vec{r}_4)$.
Next, diagrams (a) and (d) are obtained from (b) and (c)
by a complex conjugation. Finally, taking the sum of all
the four diagrams, performing angular integrals, and introducing
$x_1 = k_0 r_1$ and $x_2 = k_0 r_4$ as
new variables of integration, we get 
\begin{eqnarray}
V = V^{(a)} + \ldots + V^{(d)} =
\frac{\ell}{16 \pi k_0} F(k_0 \ell_{\varepsilon}, k_0 a),
\label{v}
\end{eqnarray}
where
\begin{eqnarray}
F(\alpha, \beta) &=& \frac{1}{\pi^{3/2}} \frac{1}{\alpha}
\int_{\beta}^{\infty} d x_1
\int_{\beta}^{\infty} d x_2
\frac{\sin(2 x_1)}{x_1} \frac{\sin(2 x_2)}{x_2}
\nonumber \\
&\times& \left\{ \exp\left[-\left(\frac{x_1 - x_2}{\alpha}
\right)^2 \right] -
\exp\left[-\left(\frac{x_1 + x_2}{\alpha}
\right)^2 \right] \right\}.
\label{FF}
\end{eqnarray}

The rest of the calculation is straightforward. The diagram of
Fig.\ \ref{fig1} yields 
$\left< \delta I(\vec{r}, t)
\delta I(\vec{r}+\Delta \vec{r}, t+\tau) \right> =
V  L(\vec{r}_0, \vec{r}) L(\vec{r}_0, \vec{r}+\Delta \vec{r})
[\ell/(4 \pi)]^2$, where the factor $[\ell/(4 \pi)]^2$ accounts for the two
vertices connecting the diffusion ladders to the points
$\vec{r}$ and $\vec{r} + \Delta \vec{r}$. As the average intensity
is $\left< I(\vec{r}, t) \right> =
[\ell/(4 \pi)]^2  L(\vec{r}_0, \vec{r})$,\cite{shapiro86}
the contribution of the diagram of Fig.\ \ref{fig1} to the
correlation function defined by Eq.\ (\ref{corr}) becomes
\begin{eqnarray}
C_0(\vec{r}, \Delta \vec{r}, 0) =
\frac{\pi}{k_0 \ell} F(k_0 \ell_{\varepsilon}, k_0 a).
\label{c01}
\end{eqnarray}
The result (\ref{c0}) is recovered for
$\ell_{\varepsilon} = a = 0$, as $F(0, 0) \equiv 1$.
Hence, the factor
$F(k_0 \ell_{\varepsilon}, k_0 a)$ describes the effects of source
size $a$ and disorder correlation length $\ell_{\varepsilon}$
on the $C_0$ correlation. In general,
$F(k_0 \ell_{\varepsilon}, k_0 a)$ is a rather complicated
function of its arguments. 
Below we
consider separately the following two cases: ({\em i\/})
$a \ll \ell_{\varepsilon} \ll \lambda$
and ({\em ii\/}) $\ell_{\varepsilon} \ll a, \lambda$.

({\em i\/}).
If $a \ll \ell_{\varepsilon} \ll \lambda$, one can
set the lower limits of integrations in
Eq.\ (\ref{FF}) equal to zero.
Then Eq.\ (\ref{FF}) can be simplified, although
the simplification requires some care, as 
the integrand of Eq.\ (\ref{FF}) does not have finite derivatives
at $\alpha = 0$ and cannot be therefore expanded in 
a power series near this point.
Instead, we represent the expression in the curved brackets of
Eq.\ (\ref{FF}) as
$2 \exp[-(x_1^2 + x_2^2)/\alpha^2] \sinh(2 x_1 x_2 /\alpha^2)$
and then expand $\sinh(\cdots)$ in Taylor series.
Carrying out integrations, we obtain\cite{grad61}
\begin{eqnarray}
F(\alpha, 0) = \frac{2}{\pi} \alpha \sum_{n=0}^{\infty}
\frac{\Gamma(n+1)}{\Gamma(n+3/2)}
{_1F_1}^2(n+1; 3/2; -\alpha^2),
\label{FF2}
\end{eqnarray}
where $_1F_1(\cdots)$ is the confluent hypergeometric function.
While the series in Eq.\ (\ref{FF2}) converges sufficiently 
fast for $\alpha > 1$, the convergence slows down as $\alpha$
becomes less than unity.

As we consider $\ell_{\varepsilon} \ll \lambda$,
we are only interested in $F(\alpha, 0)$ for
$\alpha \ll 1$. Although Eq.\ (\ref{FF2}) can be evaluated numerically
even for $\alpha < 1$, this equation is of little help if one wants
to get some qualitative insight into the small-$\alpha$ behavior
of $F(\alpha, 0)$.
To get such an insight,
we introduce $t_{1,2} = x_{1,2}/\alpha$ as new variables of integration 
in Eq.\ (\ref{FF}). This yields
\begin{eqnarray}
F(\alpha, 0) &=& \frac{2}{\pi^{3/2}} \frac{1}{\alpha}
\int_{0}^{\infty} d t_1
\int_{0}^{\infty} d t_2
\frac{\sin(2 \alpha t_1)}{t_1} \frac{\sin(2 \alpha t_2)}{t_2}
\nonumber \\
&\times&\exp\left( -t_1^2 -t_2^2 \right) \sinh(2 t_1 t_2).
\label{FF3}
\end{eqnarray}
Assuming small $\alpha$, we replace $\sin(2 \alpha t_2)$ by
$2 \alpha t_2$ and
then perform the integrals over $t_2$ and $t_1$. Finally,
we keep in a final result only the terms linear in $\alpha$.
This gives
\begin{eqnarray}
F(\alpha, 0) \simeq 1 - \frac{4}{\pi^{3/2}} \alpha,
\mbox{~~~} \alpha \ll 1.
\label{FF4}
\end{eqnarray}

A plot of $F(\alpha, 0)$ obtained by a numerical
evaluation of Eq.\ (\ref{FF2}) is presented in Fig.\ \ref{fig3}
by a solid line. The decay of $F$ is linear for $\alpha < 1$,
as predicted by Eq.\ (\ref{FF4}).
The part of the plot
corresponding to $\alpha > 1$ is meaningless for a moment,
as we have assumed $\ell_{\varepsilon} \ll \lambda$.
This part of the plot will be discussed and interpreted in
the next section, where the space-time correlation function
is considered.

As follows from Eq.\ (\ref{FF4}) and Fig.\ \ref{fig3}, the
$C_0$ correlation decreases with increase of the disorder
correlation length $\ell_{\varepsilon}$, and the characteristic
scale of this decrease is set by $\lambda$.
This is physically clear as
$C_0$ originates from scattering at a distance of order $\lambda$
from the source, which is only possible if
the wave ``identifies'' its environment
as disordered already after a distance of order $\lambda$.
Such an identification implies that
$\delta \varepsilon(\vec{r}, t)$ changes significantly at the scale of
$\lambda$, or, in other words, that the correlation length
of the fluctuations of $\delta \varepsilon(\vec{r}, t)$ 
is shorter than $\lambda$. On the basis of
this reasoning, we expect that $C_0$ should vanish for
$\ell_{\varepsilon} > \lambda$, while for
$\ell_{\varepsilon} \sim \lambda$ the behavior of $C_0$
may
be sensitive to a particular model of disorder.
In contrast, the linear initial decay (\ref{FF4}) of the $C_0$ correlation
function with the disorder correlation length $\ell_{\varepsilon}$
seems to be rather general and independent
of the model of disorder. We have verified that
the linear law (\ref{FF4}) holds (with an apparent change of
numerical factors) not only for the Gaussian
covariance of $\delta \varepsilon(\vec{r}, t)$
[Eq.\ (\ref{corrfunc})], but also for the exponential one
$B_{\varepsilon}(\Delta \vec{r}) = \sigma_{\varepsilon}^2
\exp( -\Delta r/\ell_{\varepsilon})$, as well as for
$B_{\varepsilon}(\Delta \vec{r}) = \sigma_{\varepsilon}^2
\Pi(\Delta r/\ell_{\varepsilon})$, where
$\Pi(x) = 1$ for $x \leq 1$ and
$\Pi(x) = 0$ for $x > 1$.

An essential result of our analysis is that the
$C_0$ correlation function appears to be
sensitive to the correlation length $\ell_{\varepsilon}$
of the dielectric function fluctuations, which is
a {\em short-distance\/} parameter of disorder. We remind that
all the previously studied types of correlation
($C_1$, $C_2$, $C_3$) are not sensitive to such details of the
scattering medium, and only depend on the mean free path $\ell$.
We conclude then that the sensitivity to the short-distance properties
of disordered medium is a distinctive feature of the $C_0$
correlation function. Measurements of $C_0$ could thus provide
unambiguous information on the disorder correlation length.
It is rather surprising that such information is not lost as a
result of multiple scattering, as one would expect, but determines
the long-range spatial {\em correlation\/} in the multiple-scattering
speckle pattern.

Another important outcome of the presented consideration
is that the correlation
length $\ell_{\varepsilon}$
entering into Eq.\ (\ref{c01}) is that characteristic for
the disorder in the immediate neighborhood of the source (more precisely,
at a distance of order $\lambda$ from it). The information about
the rest of the medium is only contained in the mean
free path $\ell$. Measurements of the $C_0$ correlation
could therefore be used for ``diffuse microscopy'' of disordered
media with spatial resolution of order $\lambda$. A possibility
of such a microscopy stems from the enhanced sensitivity of
the multiple-scattering speckle pattern to the properties of
disorder in the vicinity of the source.

({\em ii\/}). In the case  of $\ell_{\varepsilon} \ll a, \lambda$,
we can set $\ell_{\varepsilon} = 0$ and consider the
limit $\alpha \rightarrow 0$ of Eq.\ (\ref{FF}). The exponentials
in the curved brackets of Eq.\ (\ref{FF}) become
$\pi^{1/2} \alpha \delta(x_1 - x_2)$ and
$\pi^{1/2} \alpha \delta(x_1 + x_2)$, respectively,
which yields
\begin{eqnarray}
F(0, k_0 a) &=& 1 - \frac{2}{\pi} \left[
\mathrm{Si}(4 k_0 a) - \frac{\sin^2(2 k_0 a)}{2 k_0 a} \right]
\nonumber \\
&\simeq&\cases{1 - (4/\pi) k_0 a,
&$k_0 a \ll 1,$ \cr
1/(2 \pi k_0 a),
&$k_0 a \gg 1,$} 
\label{fb}
\end{eqnarray}
where $\mathrm{Si}(\cdots)$ is the sine integral, and the second line of the
equation shows simplified versions of the first one in
two limiting cases.
Eq.\ (\ref{fb}) is plotted in Fig.\ \ref{fig3} by a dashed line.
As $k_0 a \ll 1$, the decay of $F(0, k_0 a)$ is linear, similarly to that
of $F(k_0 \ell_{\varepsilon}, 0)$ (see the solid line in Fig.\ \ref{fig3}).
When $k_0 a$ becomes larger than unity,
$F(0, k_0 a)$ decreases as $(k_0 a)^{-1}$.
This indicates that although the small size of the source is important
for the magnitude of the $C_0$ correlation,
the correlation process leading to $C_0$
survives even for extended ($a \neq 0$) sources.

The above analysis holds for $\Delta r > \ell$. If
$\Delta r < \ell$, an additional contribution
to the $C_0$ correlation function arises due to scattering in
the neighborhood of the observation points.
In a complete analogy with the term given by Eq.\ (\ref{c01}),
this second contribution is determined by the short-distance
properties of disorder in the vicinity of the observation points
$\vec{r}$ and $\vec{r} + \Delta \vec{r}$.
For $\Delta r = 0$, the contributions owing to
scattering near the source and near the observation point
become equal and their sum determines a correction
to the second moment of intensity fluctuations
$\left< \delta I(\vec{r}, t)^2 \right>$.
Hence, the most of the conclusions drawn
concerning the $C_0$ correlation are also true
(with obvious insignificant modifications)
for $\left< \delta I(\vec{r}, t)^2 \right>$.

\section{Space-time correlation}
\label{stemporal}

We now turn to the case of the space-time correlation function,
allowing $\tau \neq 0$ in Eq.\ (\ref{corr}). It is obvious from
general considerations that the correlation function of
intensity of a multiple-scattered wave should
decrease with $\tau$ once the disordered medium is subject to some
non-periodic dynamics. This decrease was first discovered for
the $C_1$ correlation function,\cite{maret87,stephen88}
and recently has been
also studied for $C_2$ (see Ref.\ \onlinecite{frank97}) and $C_3$
(see Ref.\ \onlinecite{frank98}).

As the role of the disorder correlation length $\ell_{\varepsilon}$
has been studied in the previous section,
and in order to emphasize the effect of the medium {\em dynamics,}
we assume $\ell_{\varepsilon} = 0$ in the present section.
For the sake of concreteness, we consider the disordered medium
as a suspension of point-like scatterers which diffuse with a diffusion
constant $D_B$. The relevant correlation function of the
dielectric constant fluctuations is then\cite{stephen88,golub84}
\begin{eqnarray}
B_{\varepsilon}(\Delta \vec{r}, \tau) =
\left< \delta \varepsilon(\vec{r}, t) \delta
\varepsilon(\vec{r}+\Delta \vec{r}, t+\tau) \right> =
\frac{u}{(4 \pi D_B \tau)^{3/2}}
\exp\left( -\frac{\Delta r^2}{4 D_B \tau} \right).
\label{corrfunct}
\end{eqnarray}

Calculation of the diagrams of Figs.\ \ref{fig1}, \ref{fig2} is performed very
similar to the case of $\tau = 0$. We finally obtain
\begin{eqnarray}
C_0(\vec{r}, \Delta \vec{r}, \tau) =
\frac{\pi}{k_0 \ell} F\left(\sqrt{\tau/\tau_0}, k_0 a \right),
\label{c02}
\end{eqnarray}
where $F(\cdots)$ is given by Eq.\ (\ref{FF}),
and $\tau_0 = (4 k_0^2 D_B)^{-1}$ is the single-scattering
correlation time.\cite{maret87,stephen88} 
To simplify the further analysis, we assume $a \ll \lambda,
(\tau D_B)^{1/2}$
which allows us to extend the lower limits of integrations
in Eq.\ (\ref{FF}) to zero, and to reduce Eq.\ (\ref{FF}) to
Eq.\ (\ref{FF2}).

An approximate expression for $F(\alpha, 0)$ in the limit of small
$\alpha$ has been found in the previous section
[see Eq.\ (\ref{FF4})]. To analyze the behavior of $F(\alpha, 0)$
for $\alpha \gg 1$, we
take the large-$\alpha$ limit of Eqs.\ (\ref{FF}), (\ref{FF2}):
\begin{eqnarray}
F(\alpha, 0) \simeq \frac{1}{\pi^{3/2}} \frac{1}{\alpha^3},
\mbox{~~~} \alpha \gg 1.
\label{FF5}
\end{eqnarray}
Hence, the asymptotic behavior of the space-time correlation function
can be summarized as follows:
\begin{eqnarray}
C_0(\vec{r}, \Delta \vec{r}, \tau) \simeq
\frac{\pi}{k_0 \ell} \times
\cases{1 - (4/\pi^{3/2}) \sqrt{\tau/\tau_0},
&$\tau \ll \tau_0$, \cr
(1/\pi^{3/2}) (\tau/\tau_0)^{-3/2},
&$\tau \gg \tau_0$.}
\label{c03}
\end{eqnarray}
These approximate results are well supported by the numerical
evaluation of Eq.\ (\ref{FF2}) which yields the solid line
of Fig.\ \ref{fig3}. Linear initial decay of $F$ (and, consequently,
of $C_0$) with $\sqrt{\tau/\tau_0}$ is followed by a power-law
long-time tail, as predicted by Eq.\ (\ref{FF5}).

It is instructive to compare the initial decay of the $C_0$
correlation function with
that of $C_1$, which for the considered case of a point source
in an infinite medium and for $\Delta r = 0$ reads\cite{stephen88}
\begin{eqnarray}
C_1(\vec{r}, 0, \tau) \simeq
1 -  2 \frac{\left| \vec{r} - \vec{r}_0 \right|}{\ell}
\sqrt{\frac{3 \tau}{2 \tau_0}},
\mbox{~~~} \tau \ll \tau_0
\left( \frac{\ell}{\left| \vec{r} - \vec{r}_0 \right|} \right)^2.
\label{c1}
\end{eqnarray}
For $\Delta r \neq 0$, $C_1$ decreases exponentially with
$\Delta r / \ell$ and thus becomes negligible for
$\Delta r > \ell$.\cite{shapiro86}
Although the functional forms of $C_0(\vec{r}, \Delta \vec{r}, \tau)$
and $C_1(\vec{r}, 0, \tau)$ for short
$\tau$ are the same,
the typical correlation time $\tau_0$ corresponding to $C_0$
is much larger than the correlation time
$\tau_0 (\ell/\left| \vec{r} - \vec{r}_0 \right|)^2$
corresponding to $C_1$, as $\left| \vec{r} - \vec{r}_0 \right| \gg \ell$.
This stems from the fact that the decay of $C_1$ is due to
multiple scattering events, each giving a contribution
to the decorrelation process. The factor
$\left| \vec{r} - \vec{r}_0 \right|/\ell$ in
Eq.\ (\ref{c1}) reflects the increase of the effective number
of scattering events with $\left| \vec{r} - \vec{r}_0 \right|$.
In contrast, the decay of $C_0$ with $\tau$ originates from
single scattering events in the neighborhood of the source. The
following propagation of waves to the distant points $\vec{r}$
and $\vec{r} + \Delta \vec{r}$ depicted by the ladder propagators
in Fig.\ \ref{fig1}, does not lead to any additional decorrelation, and thus
the characteristic decay time of 
$C_0(\vec{r}, \Delta \vec{r}, \tau)$ is insensitive to
$\vec{r} - \vec{r}_0$ and $\vec{r} + \Delta \vec{r} - \vec{r}_0$.

It is worthwhile to note that there exists another important
difference between the $C_0$ and $C_1$ correlation functions.
Eq.\ (\ref{c1}) for $C_1$ is derived under assumption of
statistically homogeneous scatterer dynamics throughout the whole
space [the characteristic time $\tau_0$ in Eq.\ (\ref{c1}) is
assumed to be the same for all parts of the medium].
If, for instance, the scatterer motion in some region
of space differs from that in the rest of the medium,
the difference is reflected directly on the decay of $C_1$.
This property of $C_1$ allows its use for imaging of
scatterer dynamics in turbid media\cite{boas95}.
Moreover, space-resolved measurements of $C_1$ permit to distinguish
between Brownian and directed (flow) motion of scatterers
inside some bounded region of space, and to image scatterer
flows in multiple-scattering media.\cite{heck96}
In contrast, as the decay of $C_0$ is only sensitive to the motion
of scatterers in the vicinity of the source, any modifications
of scatterer dynamics in other parts of the medium do
not affect the $C_0$ correlation [the characteristic time
$\tau_0$ in Eqs.\ (\ref{c02}), (\ref{c03}) is that corresponding to
the immediate neighborhood of the source]. The $C_0$ correlation
function is not therefore suitable for imaging of the bulk of disordered
media.
As a compensation of this drawback, $C_0$ has, however,
an important advantage as compared to $C_1$. The advantage
is the spatially-selective sensitivity to the scatterer
motion in the immediate neighborhood of the source.
This sensitivity opens up a possibility of ``dynamic
diffuse microscopy'' of scatterer motion in multiple-scattering media.

The presented treatment of the space-time correlation is valid
for $\Delta r > \ell$. If $\Delta r < \ell$, an additional
contribution to the correlation function appears due to scattering
in the neighborhood of the observation points, just like in the case
of purely spatial correlation. For $\Delta r = 0$, the sum of the
two contributions yields a correction to the $C_1$
correlation function [Eq.\ (\ref{c1})].
A possible existence of such a correction has been suggested by
Scheffold {\em et al.},\cite{frank97} although they considered
a different experimental geometry.
This correction
is uniquely determined by the motion of scatterers
in the neighborhoods of the source and the observation point.
We emphasize that $C_0 \ll C_1$ for $\Delta r = 0$,
while for $\Delta r > \ell$ the $C_1$ term disappears and
the leading contribution to the correlation function is
given by $C_0$, since $C_2 \sim (k_0 \ell)^{-2} \ll C_0$.

\section{Conclusion}
\label{concl}

In the present paper, we consider a recently
suggested\cite{shapiro99} novel type of intensity correlation
in disordered media ($C_0$).
We show that the $C_0$ correlation function exhibits
a strongly {\em non-universal\/} behavior in the sense that
it is sensitive to the short-distance properties of disorder
and geometrical parameters (e.g., size) of the localized source of waves.
In particular, we find that $C_0$ depends explicitly
on the correlation length of dielectric function fluctuations
$\ell_{\varepsilon}$ and the size of the source $a$.
More precisely, the $C_0$ correlation is damped down
as either $\ell_{\varepsilon}$ or $a$ is increased.
As a function of $k_0 \ell_{\varepsilon}$, $C_0$ exhibits
linear decay for $k_0 \ell_{\varepsilon} \ll 1$.
As a function of $k_0 a$,
the $C_0$ correlation function also decays linearly
for $k_0 a \ll 1$, and
$C_0 \propto (k_0 a)^{-1}$ for $k_0 a \gg 1$.
The sensitivity of the $C_0$ correlation to the
short-distance
properties of disorder (e.g., to $\ell_{\varepsilon}$) is
its distinctive feature, since all other types of correlation
commonly considered for
waves in disordered media ($C_1$, $C_2$, $C_3$)
are insensitive to the details of scattering medium,
and only depend on the mean free path, which is an integral
parameter of disorder.

Next, we calculate the space-time correlation function
of intensity fluctuations
$C_0(\vec{r}, \Delta \vec{r}, \tau)$.
We show that this correlation function is only sensitive
to the scatterer motion in the immediate neighborhood of the source,
while being insensitive to the dynamics of the rest of disordered
medium. $C_0(\vec{r}, \Delta \vec{r}, \tau)$ is independent
of $\vec{r}$, $\Delta \vec{r}$ and decays linearly as a
function of $\sqrt{\tau/\tau_0}$ for $\tau \ll \tau_0$
($\tau_0$ is the single-scattering correlation time).
For large $\tau$ ($\tau \gg \tau_0$), the decays of $C_0$
is power-law: $C_0 \propto (\tau/\tau_0)^{-3/2}$. 

Finally, we suggest that the spatially-selective sensitivity
of the $C_0$ correlation function to the properties and dynamics
of disordered medium in the neighborhood of the source
(and detector for $\Delta r < \ell$) may be used to develop a
sort of ``diffuse microscopy'' of multiple-scattering media.

\acknowledgments
We would like to thank B. A. van Tiggelen for
many stimulating discussions.

\newpage
\begin{center}
\bf FIGURE CAPTIONS
\end{center}

\begin{figure}
\caption{The leading diagram contributing to the correlation
function $\left< \delta I(\vec{r}, t)
\delta I(\vec{r}+\Delta \vec{r}, t+\tau) \right>$.
$L$'s denote the ladder propagators,
the left propagator corresponds to time $t$,
and the right one --- to $t + \tau$.
Solid lines with arrows denote retarded (with arrows
directed from $\vec{r}_0$ to $\vec{r}$ or
$\vec{r} + \Delta \vec{r}$) and advanced (with arrows
directed from $\vec{r}$ or
$\vec{r} + \Delta \vec{r}$ to $\vec{r}_0$)
average Green functions, $\overline{G}$ and
$\overline{G}^*$, respectively.
$V$ is the vertex located at $\vec{r}_0$. Diagrams
contributing to $V$ are shown in Fig.\ 2.}
\label{fig1}
\end{figure}

\begin{figure}
\caption{The diagrams contributing to the vertex $V$.
Solid lines with arrows directed from $\vec{r}_0$
to some other point denote retarded average Green functions
$\overline{G}$,
solid lines with arrows directed to $\vec{r}_0$
denote advanced average Green functions $\overline{G}^*$.
In each diagram, the left pair of Green functions corresponds to time $t$,
and the right pair --- to $t+\tau$.
Crosses indicate scattering
(a factor $k_0^2$ is associated with each cross),
dashed lines denote correlation function
$B_{\varepsilon}(\Delta \vec{r}, \tau)$ of the fluctuating part
of the dielectric constant. Integrations are assumed over 
$\vec{r}_1$, $\vec{r}_2$, $\vec{r}_3$, and $\vec{r}_4$.
$\vec{r}_0$ is the source position.}
\label{fig2}
\end{figure}

\begin{figure}
\caption{Function $F(\alpha, \beta)$ describing the effects of non-zero
correlation length $\ell_{\varepsilon}$, size of the
source $a$, and time delay $\tau$ on the $C_0$ correlation
function. The solid line is $F(\alpha, 0)$, and the dashed line
is $F(0, \beta)$.
The attenuation factor which multiplies the $C_0$ value found
for $\ell_{\varepsilon} = a = 0$ and $\tau = 0$ is obtained by setting
$\alpha = k_0 \ell_{\varepsilon}$ (disorder with non-zero
correlation length, $k_0 \ell_{\varepsilon} \ll 1$),
$\alpha = (\tau/\tau_0)^{1/2}$ (scatterers undergoing Brownian
motion), or
$\beta = k_0 a$ (non-zero size of the source).}
\label{fig3}
\end{figure}


\begin{references}
\bibitem{source}
This is one of the possible and, perhaps, the simplest
way to treat the case of the source of non-zero size.
We expect that the final result is not very sensitive to
the exact form of boundary conditions at
$\left| \vec{r} - \vec{r}_0 \right| = a$. An essential
feature of our model of the source
is that no scattering occurs at $\left| \vec{r} - \vec{r}_0 \right| < a$.
\end{references}
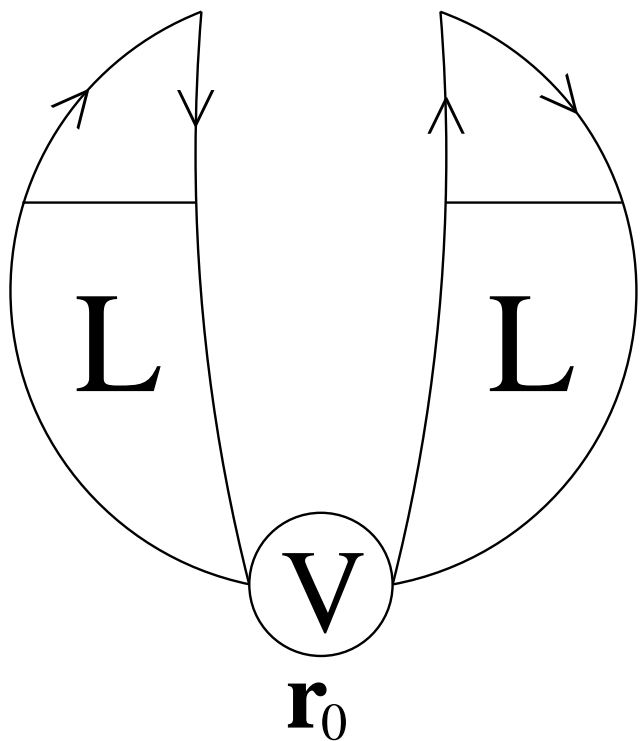
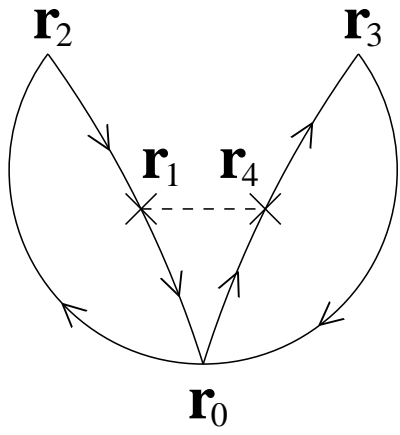
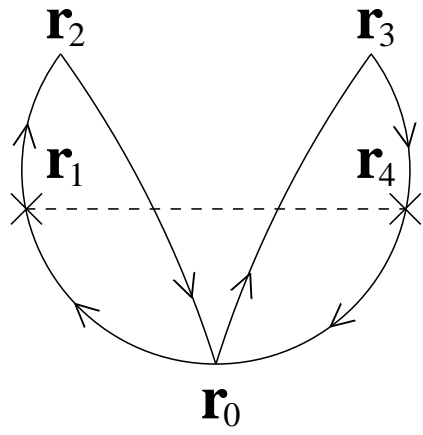
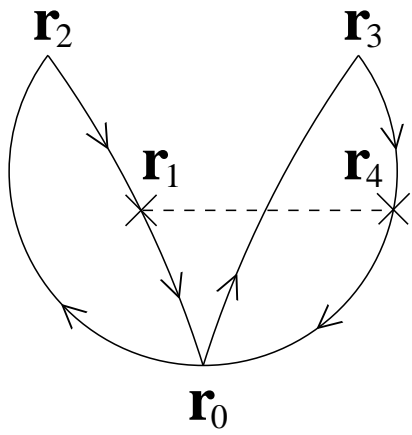
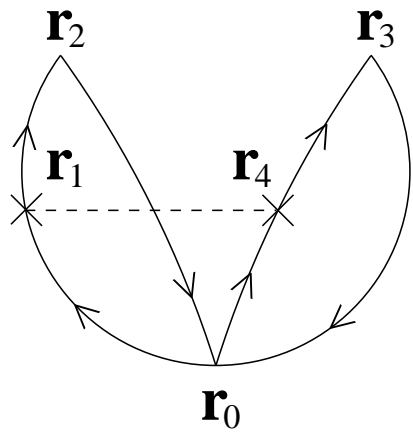
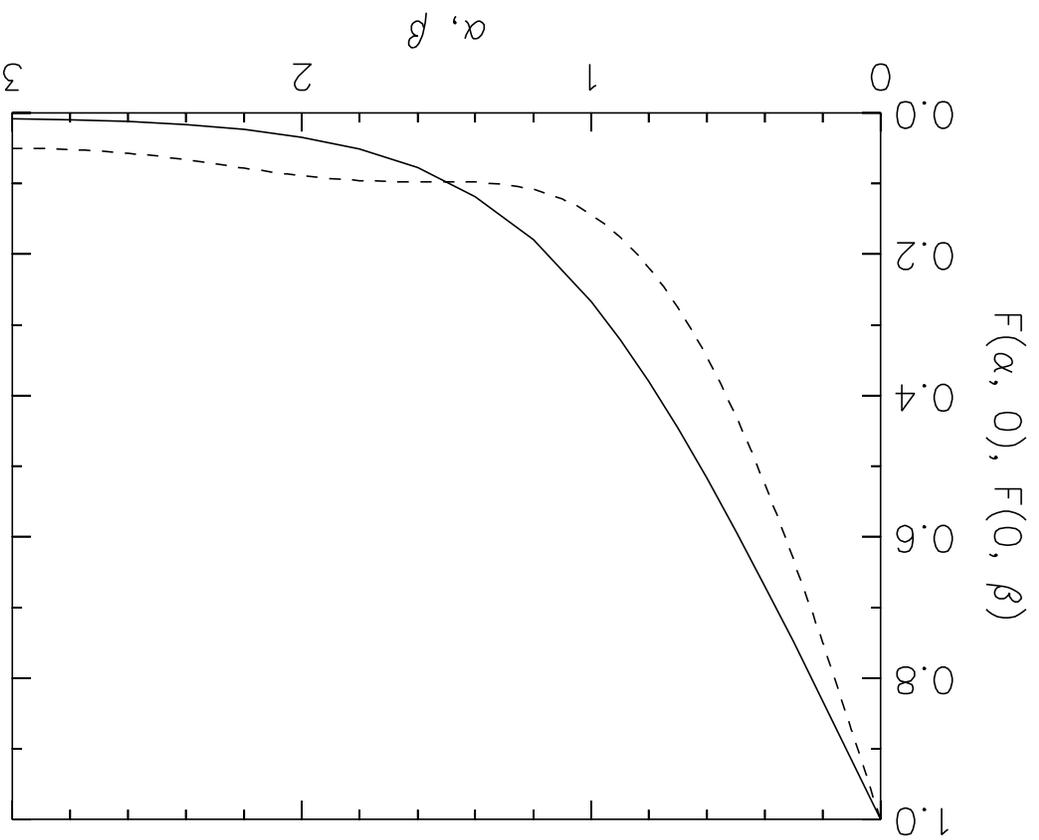
\end{document}